\documentclass[apjl]{emulateapj}
\usepackage{natbib,graphicx,apjfonts}
\def\arcdeg{\hbox{$^\circ$}}
\def\arcmin{\hbox{$^\prime$}}
\def\arcsec{\hbox{$^{\prime\prime}$}}

\def\phn{\phantom{0}}     % Phantom numeral for aligning columns in tables
\def\simgt{\lower 2pt \hbox{$\, \buildrel {\scriptstyle >}\over {\scriptstyle \sim}\,$}}
\def\simlt{\lower 2pt \hbox{$\, \buildrel {\scriptstyle <}\over {\scriptstyle \sim}\,$}}

\def\chandra{{\slshape Chandra\/}}

\def\heao1{{\slshape HEAO1\/}}

\def\spitzer{{\slshape Spitzer\/}}

\def\xray{\mbox{X-ray}}
\shorttitle{Constraints on the contribution of 24$\mu$m sources to the CXB}
\shortauthors{Steffen et al.}

\submitted{Submitted to \apj\ Letters, 2007 May 14}
\begin{document}

%% LaTeX will automatically break titles if they run longer than
%% one line. However, you may use \\ to force a line break if
%% you desire.exit

\title{{\slshape Chandra} stacking constraints on the contribution of 24 micron {\slshape Spitzer} sources to the unresolved cosmic X-ray background.}

\author{A.~T.~Steffen\altaffilmark{1},
        W.~N.~Brandt\altaffilmark{1},
        D.~M.~Alexander\altaffilmark{2},
        S.~C.~Gallagher\altaffilmark{3},
        B.~D.~Lehmer\altaffilmark{1}}
\altaffiltext{1}{Department of Astronomy and Astrophysics, 525 Davey Laboratory, Pennsylvania State University, University Park, PA 16802.}
\altaffiltext{2}{Department of Physics, University of Durham, South Road, Durham, DH1 3LE, UK.}
\altaffiltext{3}{Department of Physics and Astronomy, University of California, 430 Portola Plaza, Box 951547 Los Angeles CA, 90095-1547.}

\begin{abstract}
  We employ \xray\ stacking techniques to examine the contribution
  from \xray\ undetected, mid-infrared--selected sources to the
  unresolved, hard ($6-8$~keV) cosmic \xray\ background (CXB).  We use
  the publicly available, $24\mu$m {\slshape Spitzer Space Telescope}
  MIPS catalogs from the Great Observatories Origins Deep Survey
  (GOODS) - North and South fields, which are centered on the 2~Ms
  \chandra\ Deep Field-North and the 1~Ms \chandra\ Deep Field-South,
  to identify bright ($S_{24\mu{\rm m}} > 80\mu$Jy) mid-infrared
  sources that may be powered by heavily obscured AGNs.  We measure a
  significant stacked \xray\ signal in all of the \xray\ bands
  examined, including, for the first time, a significant ($3.2\sigma$)
  $6-8$~keV stacked \xray\ signal from an X-ray undetected source
  population.  We find that the \xray -undetected MIPS sources make up
  about $2\%$ (or less) of the total CXB below 6~keV, but about $6\%$
  in the $6-8$~keV band.  The $0.5-8$~keV stacked \xray\ spectrum is
  consistent with a hard power-law ($\Gamma = 1.44 \pm 0.07$), with
  the spectrum hardening at higher \xray\ energies.  Our findings
  show that these bright MIPS sources do contain obscured AGNs, but
  are not the primary source of the unresolved $50\%$ of $6-8$~keV
  CXB.  Our study rules out obscured, luminous QSOs as a significant
  source of the remaining unresolved CXB and suggests that it most
  likely arises from a large population of obscured, high-redshift ($z
  \simgt 1$), Seyfert-luminosity AGNs.
  
\end{abstract}
%% Keywords should appear after the \end{abstract} command. The uncommented
%% example has been keyed in ApJ style. See the instructions to authors
%% for the journal to which you are submitting your paper to determine
%% what keyword punctuation is appropriate.


\section{Introduction}
Deep ``blank-sky'' surveys with the {\em Chandra \xray\ Observatory}
(hereafter \chandra ) have resolved $ \sim50 - 90\% $ of the hard ($
2-8 $ keV) cosmic \xray\ background (CXB) into discrete sources
\citep[e.g.,][]{bauer04,de_luca04,hickox06}, with the resolved
fraction decreasing with increasing \xray\ energy
\citep{worsley04a,worsley05}.  Subsequent spectroscopic observations
have revealed that the majority of the sources are Active Galactic Nuclei \citep[AGNs;
e.g.,][]{barger03b,steffen04,szokoly04}.  This is consistent with CXB
synthesis models, which rely on the assumptions of the ``unified'' AGN
model \citep[see][for a review]{antonucci93}, that predict the
observed power-law ($\Gamma \simeq 1.4$) shape of the CXB is created
by the integrated \xray\ emission from both soft ($\Gamma \sim 1.8$),
unobscured and harder ($\Gamma < 1.4$), obscured\footnote{The term
  `obscured' refers to \xray\ obscuration, unless otherwise noted.}
AGNs \citep[e.g.,][]{comastri95,gilli07}.

Although deep \xray\ surveys find the highest sky density of AGNs to
date \citep[$\sim 7000$ deg$^{-2}$;][]{bauer04}, the decreasing
resolved fraction of the CXB with energy suggests that there exists an
additional, highly-obscured AGN population that is missed in even the
deepest \xray\ surveys \citep{worsley05}.  While the majority of the
\xray\ emission from these sources is attenuated, a small fraction of
hard \xray\ photons can penetrate the obscuring torus.  In addition,
hard (rest-frame $>10$~keV) \xray\ emission can be scattered into the observer's
line-of-sight via Compton reflection, which comprises a relatively
larger fraction of the total observed \xray\ emission from a heavily
obscured AGN.  The fraction of hard \xray\ photons emitted from a
heavily obscured source would likely be too small to identify the
source individually, but the hard \xray\ emission from many of these
undetected, Compton-thick sources could comprise the unresolved CXB.

Since the absorbed energy in obscured AGNs is re-emitted in the mid-
and far-infrared, obscured AGNs should be bright \spitzer\ sources.
The dust reprocessing of AGN accretion energy heats the dust to higher
temperatures than can be achieved by heating via stellar processes,
and this emission appears to be relatively isotropic \citep{lutz04}.
The strong rest-frame $3-8 \mu$m thermal continuum produced by the
AGN-heated dust can be used to separate luminous AGNs from starbursts
using infrared color-color selection \citep{laurent00,lacy04,stern05}
and infrared power-law selection \citep{alonso-herrero06,donley07a}.
Once candidate AGNs have been identified, \xray\ stacking analyses can
be used to measure their average X-ray properties and determine their
contribution to the CXB below the detection limit of individual X-ray
sources.

In this Letter, we use \xray\ stacking techniques to examine the
contribution of MIPS-detected GOODS sources to the unresolved
component of the CXB.  In \S2 we describe the \xray\ and infrared data
used in this study.  Our analyses are presented in \S3.  Our
discussion and conclusions are in \S4.  We use the cosmological
parameters $H_{0} = 70$ km s$^{-1}$ Mpc$^{-1}$, $\Omega_{M} = 0.3$,
and $\Omega_{\Lambda} = 0.7$.

\section{Sample}

The 2 Ms \chandra\ Deep Field-North and 1 Ms \chandra\ Deep
Field-South are currently the deepest $0.5-8$~keV surveys
\citep{giacconi02,alexander03}.  The high spatial resolution of
\chandra\ ($\sim 0 \farcs 5$ at the aimpoint) and \chandra 's low
background make these images ideal for examining the resolved fraction
of the CXB.  We use the \xray\ catalogs of \citet{alexander03} for
both the \mbox{CDF-N} and \mbox{CDF-S}, which contain 503 and 326 \xray\ sources,
respectively, detected using {\ttfamily wavdetect} at a significance
threshold of $10^{-7}$.  We also consider the supplementary \xray\
source catalogs for both fields, which add an additional 79 (42)
sources to the CDF-N (CDF-S); these sources are detected with a looser
threshold ($10^{-5}$), but because they are coincident with optically
bright ($R<23$) sources, the vast majority ($>90\%$) are likely real.

%
% FIGURE 1
%
\begin{figure}[t]
\includegraphics[angle=90,scale=0.42]{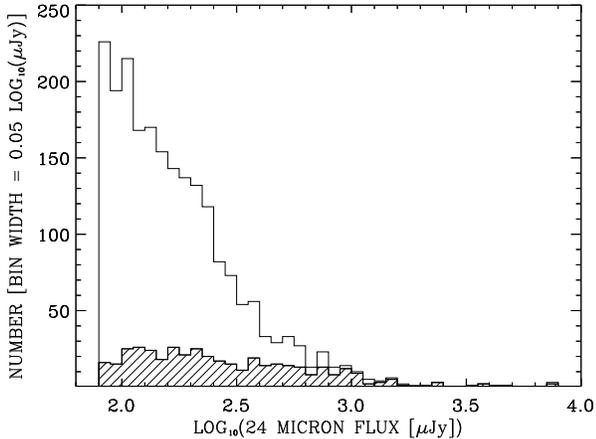}
\caption{\label{mips_xray_hist}Histogram of 24$\mu$m fluxes for the
  2147 MIPS-selected sources in the GOODS-North \& South fields
  ({\slshape open histogram}).  The 24$\mu$m flux distribution of MIPS
  sources with \xray\ counterparts is shown for comparison ({\slshape
    hatched histogram}).}
\end{figure}

To identify obscured AGN candidates, we use \spitzer\ data from the
Great Observatories Origins Deep Survey \citep[GOODS;][]{dickinson05}.
The GOODS infrared catalogs consist of observations in the four IRAC
bands \citep[3.6, 4.5, 5.8, and 8.0$\mu$m;][]{fazio04} and one MIPS
band \citep[24$\mu$m;][]{rieke04}.  The contribution to the CXB from
the IRAC-selected sources will be presented in a subsequent paper
(Steffen et al. 2007, in prep).  In this Letter we measure the
contribution of the MIPS sources to the CXB.

The GOODS MIPS observations are publicly available\footnote{Available
  at \url{http://ssc.spitzer.caltech.edu/legacy/goodshistory.html}.}
as part of GOODS data releases DR1 (GOODS-North) and DR3
(GOODS-South).  The GOODS-North and GOODS-South MIPS-selected catalogs
contain 1199 and 948 sources, respectively, extending down to a
$24\mu$m flux limit of \mbox{$\sim80\mu$Jy.}  We cross-correlated the
MIPS source positions with the aforementioned \xray\ catalogs using a
$3\arcsec$ matching radius.  We found 253 (156) CDF-N (CDF-S) MIPS
sources matched to an \xray\ counterpart within $3\arcsec$ (with an
expectation of $\sim20$ spurious matches).   In
Figure~\ref{mips_xray_hist}, we present histograms showing the
distribution of 24$\mu$m fluxes for all MIPS-selected GOODS
sources ({\slshape open histogram}) and for the MIPS sources with
\xray\ counterparts ({\slshape hatched histogram}).  It is apparent
that while the number of MIPS sources increases at fainter
fluxes, the number of sources with \xray\ counterparts increases 
slowly, so the fraction of MIPS sources with a detected
\xray\ counterpart decreases as one examines fainter 24$\mu$m MIPS
sources.  This is consistent with the idea that luminous AGNs power
the brightest 24$\mu$m sources, while the fainter sources can be
powered by the more common dusty starbursts.

\section{Analysis}
%
% FIGURE 2
%
\begin{figure}[b]
\epsscale{1.0}
\plotone{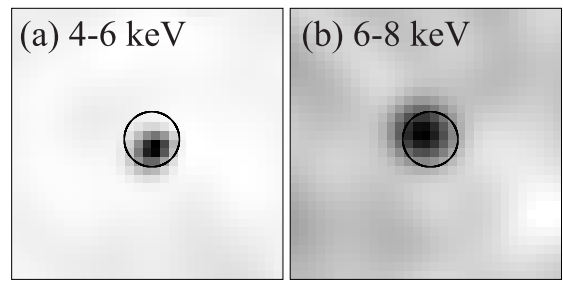}
\caption{\label{stacked_thumb}({\slshape a}) Stacked, adaptively
  smoothed, $4-6$ keV \xray\ thumbnail for the $638$ MIPS-selected AGN
  candidates.  Adaptive smoothing was performed using the
  \texttt{csmooth} algorithm in CIAO.  The $3\arcsec$ diameter aperture
  used for source photometry is shown.  ({\slshape b}) Stacked,
  adaptively smoothed $6-8$ keV \xray\ thumbnail.}
\end{figure}

%
% TABLE 2
%
\begin{deluxetable*}{ccccccccc}[t]
  \tablewidth{0pt}
  \tablecolumns{9}
  \tablecaption{\label{mips_stacking_results}Results of \xray\ Stacking Analysis of MIPS Sources}
  \tablehead{
    \colhead{Energy Band} &
    \colhead{Total Counts} &
    \colhead{Mean Background\tablenotemark{a}} & 
    \colhead{Exposure Map} &
    \colhead{Encircled Energy} &
    \colhead{Area\tablenotemark{c}} &
    \colhead{Total Intensity} &
    \colhead{\% of CXB\tablenotemark{d}} &
    \colhead{S/N} \\
    \colhead{[keV]} &
    \colhead{[counts]} &
    \colhead{[counts]} &
    \colhead{(mean) [cm$^{2}$ s]} &
    \colhead{Fraction\tablenotemark{b}} &
    \colhead{[deg$^{2}$]} &
    \colhead{[ergs cm$^{-2}$ s$^{-1}$ deg$^{-2}$]} &
    \colhead{} &
    \colhead{[$\sigma$]} \\
    \colhead{[1]} &
    \colhead{[2]} &
    \colhead{[3]} &
    \colhead{[4]} &
    \colhead{[5]} &
    \colhead{[6]} &
    \colhead{[7]} &
    \colhead{[8]} &
    \colhead{[9]}
  }
  \startdata
  \cutinhead{Standard \chandra\ \xray\ bands}
   $0.5-8.0$ & $6857.2 \pm 82.8$ & $4698.9 \pm 6.2$  & $3.86 \times 10^{8}$ & 0.556 & $5.676 \times 10^{-2}$  & $(5.50\pm0.21) \times 10^{-13}$ & $2.35\pm0.09$ & 26.0 \\
   $0.5-2.0$ & $2904.6 \pm 53.9$ & $1429.8 \pm 3.4$  & $3.85 \times 10^{8}$ & 0.614 & $5.930 \times 10^{-2}$  & $(1.64\pm0.06) \times 10^{-13}$ & $2.16\pm0.08$ & 27.3 \\
   $2.0-8.0$ & $3952.6 \pm 62.9$ & $3269.1 \pm 5.2$  & $3.95 \times 10^{8}$ & 0.542 & $5.559 \times 10^{-2}$  & $(4.22\pm0.34) \times 10^{-13}$ & $2.43\pm0.20$ & 10.8 \\[-5pt]

   \cutinhead{Narrow \xray\ bands}
   $0.5-1.0$ & $1202.3 \pm 34.7$ & $\phn679.2 \pm 2.4$ & $2.50 \times 10^{8}$ & 0.624 & $5.950 \times 10^{-2}$  & $(0.60\pm0.04) \times 10^{-13}$ & $1.99\pm0.13$ & 15.0 \\
   $1.0-2.0$ & $1702.4 \pm 41.3$ & $\phn750.5 \pm 2.5$ & $6.94 \times 10^{8}$ & 0.610 & $5.920 \times 10^{-2}$  & $(0.85\pm0.04) \times 10^{-13}$ & $1.86\pm0.09$ & 23.0 \\
   $2.0-4.0$ & $1491.7 \pm 38.6$ & $1088.5 \pm 3.0$    & $4.10 \times 10^{8}$ & 0.578 & $5.826 \times 10^{-2}$  & $(1.36\pm0.13) \times 10^{-13}$ & $1.96\pm0.19$ & 10.4 \\
   $4.0-6.0$ & $1089.9 \pm 33.0$ & $\phn927.8 \pm 2.8$ & $4.32 \times 10^{8}$ & 0.544 & $5.588 \times 10^{-2}$  & $(0.97\pm0.20) \times 10^{-13}$ & $1.73\pm0.36$ & \phn4.9 \\
   $6.0-8.0$ & $1371.0 \pm 37.0$ & $1252.3 \pm 3.2$    & $1.69 \times 10^{8}$ & 0.504 & $5.463 \times 10^{-2}$  & $(2.83\pm0.89) \times 10^{-13}$ & $5.79\pm1.82$ & \phn3.2 \\[-8pt]
      \enddata
   \tablenotetext{a}{The small error bars for the mean background result from the much larger area used to calculate this value, relative to the size of the extraction aperture.} 
   \tablenotetext{b}{This is the mean encircled-energy fraction weighted by the exposure map values of the sources.}
   \tablenotetext{c}{This is the total area within $6\arcmin$ of the CDF-N and -S aimpoints, corrected for the masked areas within $2\times 90\%$ encircled energy of the known X-ray sources.}
   \tablenotetext{d}{Assuming the CXB normalization of \citet{hickox06}.  The CXB fraction is $11\%$ higher assuming the normalization value measured by \citet{revnivtsev05} and $6\%$ lower using the value from \citet{de_luca04}.}
\end{deluxetable*}

We used \xray\ stacking techniques to examine the contribution of
MIPS-selected AGN candidates to the unresolved CXB.  We calculated the
total number of photons within a $3\arcsec$ diamater circular
aperture, correcting for the fractional contribution from pixels only
partially covered by the aperture.  The local background is measured
by extracting the total number of counts within a $30 {\rm ~pixel}
\times 30 {\rm ~pixel}$ ($\sim 15\arcsec \times15\arcsec$) box,
centered on a source and masking all of the known X-ray and MIPS
source positions.

To calculate the expected background in the desired aperture, the
total number of background counts is scaled by the ratio of the summed
exposure map values within the source aperture and the summed exposure
map values in the local background region.  This is similar to scaling
by the relative sizes of the source aperture and background region,
except this method accounts for variations in sensitivity due to chip
gaps and edges, and HRMA mirror effects.  Our simulations have shown
that this method is both much faster and more accurate than measuring
the local background using thousands of randomly placed apertures, as
would be adopted in a Monte-Carlo approach.

For each source, the encircled-energy fraction was calculated using a
tabular parameterization of the \chandra\ PSF, provided by the CXC,
that gives the radius of a circular aperture for a given off-axis
angle, photon energy, encircled-energy fraction, and azimuthal
angle.\footnote{Available at
  \url{http://cxc.harvard.edu/cal/Hrma/psf/ECF/hrmaD1996-12-20hrci\_ecf\_N0002.fits}.}
Using our desired aperture size, mean photon energy, and the sources'
off-axis angles from the exposure-weighted mean aimpoints given by
\citet{alexander03}, we interpolated the HRMA FITS table to obtain the
encircled-energy fraction for each source.  Since the individual
images used in the CDF-N and CDF-S mosaics were taken at a variety of
roll angles, we removed the small azimuthal dependence by averaging
the PSF parameterization over all azimuthal angles.

To calculate the total \xray\ flux for all of the stacked sources, the
total background-subtracted source counts (in photons) was divided by
the sum of the mean exposure-map value within each extraction aperture
(in cm$^{2}$ s) to obtain the total photon flux for the stacked
sources.  This value was divided by the mean encircled-energy
fraction, weighted by the mean exposure-map value for each source, and
converted to energy flux by assuming a power-law spectrum with $\Gamma
= 1.4$ and correcting for Galactic absorption using the \xray\ opacity
table of \citet{morrison83} and the exposure-weighted mean Galactic
column density (N$_{\rm HI}=1.16 \times 10^{20}$~cm$^{-2}$).

In Table~\ref{mips_stacking_results}, we show the stacking results for
the 638 MIPS sources that lay within $6\arcmin$ of the
exposure-weighted mean aimpoints of the CDF-N or the CDF-S, and were
outside $2\times$ the radius of the 90\% encircled-energy fraction
of the \xray\ sources in the \citet{alexander03} main and
supplementary catalogs to avoid contamination from known \xray\
sources.  The \xray\ stacking was performed in the three standard
\chandra\ bands, the full ($0.5-8.0$ keV), soft ($0.5-2$~keV), and
hard ($2-8$~keV) bands, as well as five narrower, non-overlapping
\xray\ bands.

From Table~\ref{mips_stacking_results}, the \xray\
undetected MIPS source population makes up about 2\% of the total CXB
intensity below 6~keV.  The resolved fraction of the CXB increases to
$\sim6\%$ for the hardest \xray\ energy band analyzed here.  
This is the first time that a statistically significant, stacked
$6-8$~keV signal has been found for an \xray\ undetected population of
sources.  \xray\ stacking analyses were used to measure the
contribution of the optical GOODS sources to the CXB, but no
significant $6-8$~keV signal was found \citep{worsley06}.  To verify
the authenticity of the $3.2\sigma$ detection in the $6-8$~keV energy
band, we extracted thumbnail \xray\ images centered on each \xray\
undetected MIPS source and coadded them.  In
Figure~\ref{stacked_thumb}, we show the resulting $15\arcsec \times
15\arcsec$ stacked \xray\ thumbnail for both the ({\slshape a})
$4-6$~keV and ({\slshape b}) $6-8$~keV bands with our $3\arcsec$
diameter aperture overlaid ({\slshape black circle}).  It is clear
from these smoothed images that there is a significant stacked \xray\
signal within our extraction aperture for both \xray\ bands
($p=6.9\times10^{-4}$ in the $6-8$~keV band, assuming a one-tailed
Gaussian distribution).  Our $3\arcsec$ aperture was chosen to
maximize the stacked signal within the aperture while minimizing the
background signal.  While the peak of the stacked X-ray image is not
precisely centered within our aperture, using a larger aperture does
not increase the measured S/N for the stacked $6-8$~keV signal.

%
% FIGURE 3
%
\begin{figure}[b]
\epsscale{1.0}
\plotone{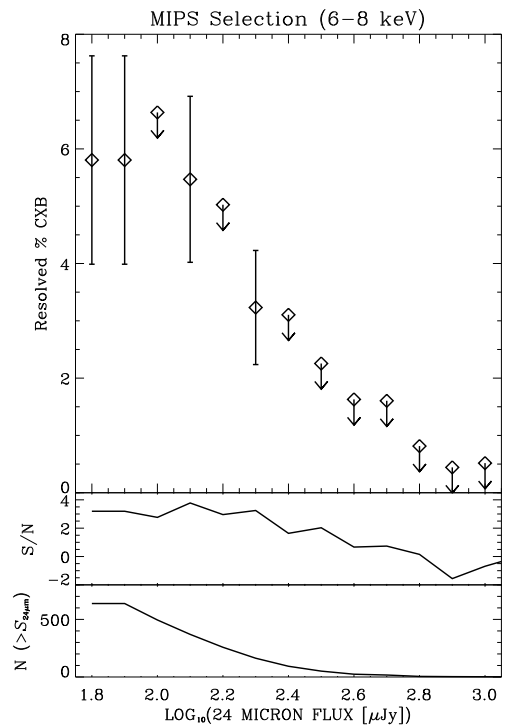}
\caption{\label{stack_vs_mag}The total resolved $6-8$~keV CXB
  fraction, the signal-to-noise of the stacked \xray\ signal, and the
  total number of stacked sources as a function of 24 $\mu$m flux
  assuming the CXB normalization of \citet{hickox06}. Error bars
  ($1\sigma$) are shown for stacked sources that have S/N > 3, and
  $3\sigma$ upper limits are given for the stacked sources that are
  not significantly detected.}
\end{figure}

While we detect a significant $6-8$~keV \xray\ signal for the stacked,
\xray --undetected MIPS population we do not know how this low-level
\xray\ flux is distributed among the sources.  Is the stacked
$6-8$~keV signal dominated by a small number of bright $24\mu$m
sources, as one might suspect from Figure~\ref{mips_xray_hist}?  To
examine the importance of infrared flux on our \xray\ stacking
analysis, we break down our earlier stacking results by performing
stacking as a function of limiting $24\mu$m flux.  In
Figure~\ref{stack_vs_mag}, we show the resolved fraction of the CXB,
the signal-to-noise (S/N), and the number of sources stacked as a
function of limiting $24\mu$m flux.  Overall, it appears that the S/N
of our stacked \xray\ signal does gradually increase as we include the
\xray\ flux from fainter MIPS sources, which suggests that the stacked
\xray\ signal is not dominated by a small number of bright $24\mu$m
sources.

\section{Discussion and Conclusions}

We present in this Letter results from an \xray\ stacking analysis of
the \xray--undetected $24\mu$m {\slshape Spitzer} MIPS sources using
the GOODS catalogs and the {\chandra} Deep Fields.  We find, for the
first time, a significant stacked \xray\ signal ($3.2\sigma$) in the
$6-8$~keV band for an \xray -undetected, AGN-candidate sample,
suggesting that at least some of these sources harbor heavily obscured
AGNs.  Approximately $12\%$ of the unresolved fraction \citep[$\sim
50\%$;][]{worsley05} of the $6-8$~keV CXB can be attributed to these
sources.  The stacked $0.5-8$~keV spectrum has a power-law photon
index of $\Gamma=1.44 \pm 0.07$, consistent with the slope of the CXB
\citep{hickox06}.  The slope of the stacked \xray\ spectrum hardens
with increasing \xray\ energy, suggesting that the flux from obscured
AGNs dominates the stacked spectrum at high \xray\ energies.

\begin{figure}[t]
\epsscale{1.0}
\plotone{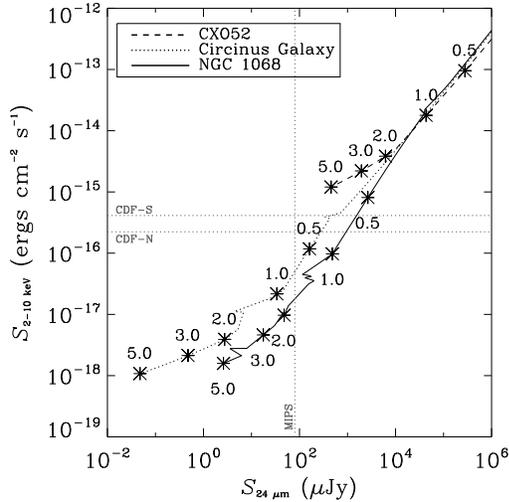}
\caption{\label{sey2_evol}The redshift evolution of the observed-frame
  $2-10$~keV flux and $24\mu$m flux density for three obscured AGNs,
  CXO52 \citep[{\slshape dashed curve}]{stern02}, NGC 1068 ({\slshape
    solid curve}), and the Circinus galaxy ({\slshape dotted curve}).
  The two dotted horizontal lines and one dotted vertical line denote the average flux limits for
  the 2 Ms CDF-N, 1 Ms CDF-S, and the GOODS MIPS catalogues, respectively.}
\end{figure}

We found evidence that there exist mid-IR-bright,
heavily obscured AGNs that are not individually detected in the \xray\
band, but we know little about the general properties of these
sources.  By examining how the \xray\ to $24\mu$m flux relation for
AGNs evolves with redshift, we can infer the AGN types that are
detected in this mid-IR catalog.  In Figure~\ref{sey2_evol}, we
examine the evolution of the \xray\ to $24\mu$m flux relation for
obscured AGNs using the observed SEDs of three obscured AGNs: the
luminous, Type 2 quasar CXO52 \citep[distant Compton-thin, obscured
QSO; $L_{2-10 \rm{~keV}} = 3.3 \times 10^{44}$ ergs s$^{-1}$,
intrinsic;][]{stern02}, NGC~1068 \citep[extreme Compton-thick,
luminous AGN; $L_{2-10 \rm{~keV}} \sim 10^{43} - 10^{44}$ ergs
s$^{-1}$, intrinsic;][]{cappi06}; and the Circinus Galaxy
\citep[Compton-thick, moderate-luminosity AGN; $L_{2-10 \rm{~keV}}
\sim 10^{42}$ ergs s$^{-1}$, intrinsic;][]{matt99}.  This is similar
to Figure~1 of \citet{martinez-sansigre06}, except here we utilize the
observed SEDs of known obscured AGNs instead of using theoretical
models.  The rest-frame \xray\ and IR fluxes were calculated by
interpolating the SEDs provided by \citet{stern02} for CXO52, and the
NASA/IPAC Extragalactic Database (NED) for NGC~1068 and the Circinus
Galaxy.

%
% FIGURE 4
%

From Figure~\ref{sey2_evol} it is apparent that a luminous,
significantly obscured AGN such as CXO52 would be detected in both the
GOODS MIPS catalog and the deep \xray\ surveys, even at $z>5$, and
would thus would not contribute to the unresolved portion of the CXB.
Even if this source were Compton thick, it would easily be detected at
$24\mu$m and would be included in our \xray\ stacking analysis.  Given
the small fraction of the CXB that we resolve at $6-8$~keV, the
unresolved CXB at these energies {\slshape cannot} be attributed to a
population of luminous, heavily obscured, ``Type 2'' Quasars at any
redshift (unless Compton-thick QSOs are significantly fainter at
$24\mu$m than their less obscured counterparts).  In addition, it is
apparent that heavily obscured, low-luminosity Seyfert 2s without
significant star-formation can be detected in the MIPS band at
$z<0.8$, but fall below the GOODS detection threshold at higher
redshifts.  This suggests that the unresolved $6-8$~keV CXB is not
emanating primarily from a population of low-luminosity, low-redshift
AGNs but could be from a population of low-luminosity, $z>0.8$ AGNs.
At these low $24\mu$m fluxes it is difficult to separate AGNs from
dusty starburst galaxies, which makes stacking analyses problematic
due to the increased background signal in the $6-8$~keV band from
obscured starbursts.

The average $0.5-2$~keV flux for our MIPS-selected AGNs is $S_{0.5-2
  {\rm ~keV}} \simeq 5.4\times10^{-18}$ ergs cm$^{-2}$ s$^{-1}$, a
factor of $\sim4.6$ lower than the flux limit of the 2~Ms CDF-N
\citep{alexander03}.  Deeper \xray\ observations will help to improve
significantly the S/N of the stacked \xray\ signal and will greatly improve
our measurements of the $6-8$~keV CXB from these sources.  In
addition, if it is possible to seperate better the obscured starbursts
from the infrared-selected AGN candidates, the significance of the
stacked \xray\ detections should improve.  We plan to address
additional infrared AGN selection techniques in a subsequent paper
\citep{steffen07b}.

\acknowledgements
%We thank the anonymous referee for helpful comments that improved the manuscript.
We gratefully acknowledge support from CXC grant GO4-5157A (A.~T.~S.,
W.~N.~B., B.~D.~L.), SAO grant AR6-7012X (A.~T.~S.), JPL grant 1278940
(A.~T.~S.), NASA LTSA grant NAG5-13035 (W.~N.~B.), the Royal
Society (D.~M.~A), and JPL grant 1268000 (S.~C.~G.).

 {\it Facilities:} \facility{{\slshape CXO} (ACIS)}, \facility{{\slshape Spitzer} (MIPS)}

%
%%% The following command ends your manuscript. LaTeX will ignore any text
%%% that appears after it.
\end{document}